# N-representability of two-electron densities and density matrices and the application to the few-body problem

Mats-Erik Pistol

Solid State Physics/Nanometer Structure Consortium, Box 118, Lund University,

S-221 00 Lund, Sweden

We have found a (dense) basis for the N-representable, two-electron densities, in which all N-representable two-electron densities can be expanded, using positive coefficients. The inverse problem of finding a representative wavefunction, giving a prescribed two-electron density, has also been solved. The two-electron densities are found to lie in a convex set in a vector space. We show that density matrices are more complicated objects than densities, and density matrices do not seem to lie in a convex set. An algorithm to compute the ground-state energy of a few-particle system is proposed, based on the obtained results , where the correlation is treated exactly.

71.10.-w, 71.15.Nc

The quantum many-body problem is a classic problem, which is very hard to solve and is very important. Of the two main approaches, density functional theory suffers from being based on an unknown energy functional and many-body perturbation theory suffers from being based on an electron gas and having difficulties treating highly correlated systems. Here we will give new information on density matrices and two-electron densities.

It is well-known that, the properties of a many-body system can be extracted explicitly from the two-particle density matrix, as long as, the operators involved couple at the most two-particles[1,2]. The ground state of any many-body system, whether atoms, molecules or extended systems, can be found by minimizing the total energy over the space of two-particle density matrices. The complicated many-body problem is seemingly reduced to a fairly simple problem. Unfortunately, the space of valid two-electron densities is not easily characterized and the problem of characterizing valid two-electron densities and density matrices is known as the N-representability problem. The problem has been solved in two different ways, either by direct construction of valid density matrices from wavefunctions[3] or by an algebraic method[4]. The complexity is large in both methods, as presently implemented, and at least the algebraic method is so far computationally very demanding[3,5]. The wavefunction method is very direct, but gives no information about whether density matrices and densities may or may not be added. The Schrödinger equation can be written in terms of density matrices of different orders, and using relations between these density matrices[6,7] it is possible to arrive at a final equation to be solved. This method presently suffers from approximations, but is nevertheless an intriguing approach for solving the many-body problem. These approximations are in part due to the difficulty in ensuring the N-representability of the density matrices[8]. The use of geminals or two-particle functions instead of the density in a density-functional type of approach has also been suggested[9]. The problem of N-representability is serious in this method as well.

Here we will give a simple and explicit method to generate N-representable two-electron densities. We will also solve the inverse problem of finding representative wavefunctions, given a prescribed two-particle density. The method is based on a unique set of fundamental or "basic" two-particle densities. The term "valid" will often be used instead of "N-representable". We will also treat two-particle density matrices and demonstrate that they are much more complicated than two-particle densities.

Consider a general N-particle wavefunction:

$$\Psi(x_1, x_2, ..., x_N) \qquad (1)$$

where a possible spin coordinate is included in the $x_i$. We define the two-particle density matrix as:

$$\Gamma(x_1', x_2' | x_1, x_2) = \binom{N}{2} \int \overline{\Psi(x_1', x_2', x_3..., x_N)} \Psi(x_1, x_2, x_3..., x_N) dx_3...dx_N \qquad (2)$$

and the two-particle density as:

$$C(x_1, x_2) = \binom{N}{2} \int \overline{\Psi(x_1, x_2, ..., x_N)} \Psi(x_1, x_2, ..., x_N) dx_3...dx_N \qquad (3)$$

$C(x_1, x_2)$ is clearly seen to be the diagonal of $\Gamma(x_1', x_2' | x_1, x_2)$, i. e., $C(x_1, x_2) = \Gamma(x_1, x_2 | x_1, x_2)$. The two-particle density encodes all two-particle correlation in the system.

If the Hamiltonian is:

$$H = T + U + V = \sum_i H_i + \sum_{i<j} H_{ij} + \sum_i V(x_i) \qquad (4)$$

with $H_i = -\nabla_i^2$ being the kinetic term, $H_{ij} = \dfrac{1}{|x_i - x_j|}$ representing the Coulomb repulsion between, e. g., electrons and V(x$_i$) an external potential, we find that

$$\langle H \rangle = \frac{2}{N-1} \int -\nabla_1^2 \Gamma(x_1',x_2'|x_1,x_2)\Big|_{x_1' \to x_1, x_2' \to x_2} dx_1 dx_2 + \int \frac{C(x_1,x_2)}{|x_1 - x_2|} dx_1 dx_2 + \\ + \frac{2}{N-1} \int C(x_1,x_2) V(x_1) dx_1 dx_2 \quad (5)$$

This shows the importance of $\Gamma(x_1', x_2' | x_1, x_2)$ and $C(x_1, x_2)$. Note that the Laplace operator works only on unprimed coordinates, and that after the operation is performed we make the substitution x´-> x. The first, kinetic, term can be written in several equivalent ways[10]. The problem of finding the minimum energy-state of a many-body problem is reduced to minimizing Eqn. (5) with respect to $\Gamma(x_1', x_2'| x_1, x_2)$. It is vital that the variation is performed over N-representable density matrices. If not, the energy will be lower than the experimental energies. The minimization is to be performed in the space of all N-representable density matrices, which avoids the issue of v-representability[11].

Spin can be included by using density matrices corresponding to the various spins orientations. Eqn. (5) will then become a sum of these terms. Spin-spin interaction terms should then be included in the Hamiltonian.

We will first treat two-particle densities. The idea is very simple. We choose a real function $\chi_0$, which has the value one in a small hyper-cube around the origin, V$_0$ (which we will call the support), and zero outside the hyper-cube. We then create a set of functions, $\chi_i$ by translating $\chi_0$ to different positions, e.g., in a cubic lattice. The lattice spacing is *a*, which will be used below in the discussion of density matrices. The supports of the $\chi_i$ are not allowed to overlap, but they should "touch" each other. If we are dealing with fermions, we create a new function, $\phi_i$, which is antisymmetric under permutation of any two of its arguments. This new function can simply be written as:

$$\phi_i(x_1,...,x_N) = C\sum_k \text{sgn}(P_k)P_k\chi_i(x_1,...,x_N) \tag{6}$$

where $P_k$ is a permutation of the indices and sgn($P_k$) gives the signature of this permutation. If we are dealing with bosons, we do not need the signature function since the function is then symmetric. The constant, C, is chosen to normalize $\phi_i$. It is important to note that the product of two different functions is zero:

$$\phi_i \cdot \phi_k = 0. \tag{7}$$

This is easily seen to be a consequence of the fact that the right hand side of Eqn. (6) sums over all permutations of the indices. Therefore, $\phi_i \cdot \phi_k \neq 0$ would imply that $\chi_i(x_1,...,x_N) = P_k\chi_n(x_1,...,x_N)$ for some permutation, apart from a possible sign. It then follows that:

$$\sum_m \text{sgn}(P_m)P_m\chi_i(x_1,...,x_N) = \sum_m \text{sgn}(P_m)P_mP_k\chi_n(x_1,...,x_N). \tag{8}$$

Since both sides of this equation runs over all permutations exactly once, we have:

$$\phi_i = \phi_n \tag{9}$$

apart from a possible sign. These functions, $\phi_i$, form an orthogonal basis with the proper symmetry under permutations. We will, from now on, call them proper basis functions. From now on we will specify the particles as electrons, although bosons are treated exactly the same.

From the proper basis functions, we obtain valid two-electron densities:

$$C_i(x_1,x_2) = \binom{N}{2}\int \overline{\phi_i(x_1,x_2,...,x_N)}\phi_i(x_1,x_2,...,x_N)dx_3...dx_N. \tag{10}$$

We will call $C_i(x_1, x_2)$ a basic two-electron density. Consider a normalized sum of two proper basis functions: $\varphi = \alpha_i \phi_i + \alpha_k \phi_k$ and form the two-electron density:

$$C_\varphi(x_1, x_2) = \binom{N}{2} \int \overline{(\alpha_i \phi_i + \alpha_k \phi_k)}(\alpha_i \phi_i + \alpha_k \phi_k) dx_3 ... dx_N$$

$$= |\alpha_i|^2 C_i(x_1, x_2) + |\alpha_k|^2 C_k(x_1, x_2) + \binom{N}{2} \int \overline{\alpha_i \phi_i} \alpha_k \phi_k + \overline{\alpha_k \phi_k} \alpha_i \phi_i dx_3 ... dx_N \quad (11)$$

The last term is zero, since $\phi_i \cdot \phi_k = 0$. It can be seen that each $\phi_i$ will generate a two-electron density, $C_i(x_1, x_2)$, such that any sum of this type of two-electron densities, with positive coefficients, will form a new valid two-electron density. Specifically, if:

$$\sum_i |\alpha_i|^2 = 1 \quad (12)$$

then

$$C(x_1, x_2) = \sum_i |\alpha_i|^2 C_i(x_1, x_2) \quad (13)$$

is an N-representable two-electron density. The two-electron densities define a convex set in a vector space since the expansion coefficients have to be positive. This set is convex since a $C_i$ + b $C_k$ belongs to the set if a+b =1 and a and b are positive. If we allow two-electron densities which are not normalized then the set will be a convex cone, which is helpful for visualization of the addition property.

Eqns. (12-13) have the simple but important consequences:

*(a) every normalized function within the convex set is a valid two-electron density since it can be expanded in basic two-electron densities.*

*(b) the normalized sum of two (or more) valid two-electron densities is a valid two-electron density, provided that the coefficients in the sum are positive.*

*(c) we can find a representative wavefunction corresponding to any two-electron density within the convex set.*

To get a representative wavefunction from a two-electron density we first expand the two-electron density according to Eqn. (13). We then find the wavefunctions corresponding to each basic two-electron density in the expansion and add these wavefunctions. Therefore, the representative wavefunction is:

$$\psi(x_1,...,x_N) = \sum_i \alpha_i \phi_i(x_1,...,x_N) \tag{14}$$

Note, however, that this wavefunction is not unique, since the coefficients can have any phase. It is only when we include the kinetic term, T, that the ground state wavefunction becomes uniquely determined by the two-electron density (Rosinas theorem)[2]. This is similar to the situation in density functional theory, where there is a unique wavefunction corresponding to a given density only if one knows that the wavefunction is also a minimum of a certain Hamiltonian (where the ground-state is assumed to be unique)[11]. We can see in general that the set of functions $e^{if(x_1,...,x_N)}\psi(x_1,...,x_N)$ all give the same two-electron density for arbitrary (sufficiently smooth) $f(x_1,...,x_N)$. From valid two-electron densities, we can obtain valid three- and n-electron densities via the representative wavefunctions and this can also be done directly. We have done numerical experiments and we find that a guessed two-electron density is almost never N-representable if N is greater than 2. Assume a classical situation with N electrons distributed on the real axis at integer positions. The density function is $f(x;a_1,a_2,...,a_N) = 1$ if $x = a_i$ and zero otherwise. The two-electron density is then $f(x_1;a_1,a_2,...,a_N) * f(x_2;a_1,a_2,...,a_N)$, where in addition $f(x,x) = 0$. The readers can quickly convince themselves that a randomly chosen function almost never factorizes in this way, and is therefore not N-representable. The situation is no better in the quantum regime.

The set $\phi = \{\phi_i\}$ is "dense" in the set of wavefunctions which are zero outside a finite region. With "dense" we mean that any wavefunction can be approximated arbitrarily close by a set of proper basis functions. In the standard way we need to increase the number of proper basis functions if we want a better approximation to a given function. When we increase the number of $\phi_i$'s we must decrease their support. A simple way to increase the number of basis functions is to perform a re-scaling. To be more explicit, if the set $\phi_m$ has support over a lattice, let $\phi_{m+1}$ have support over a lattice which is twice as dense and where the supports, V, have half the size. A finite size of $\phi$ naturally means that the system has a finite size. This is not a serious restriction, since the size can be made arbitrarily large.

This immediately implies that this procedure will also generate two-electron densities which are dense in the set of all N-representable two-electron densities (which are zero outside the size of the system). With dense we mean here that any valid two-electron density can be approximated arbitrarily closely by a sum of basic two-electron densities using an expansion with only positive coefficients, according to Eqn.(10). A function which is outside the convex set is not N-representable. The set of basic correlation functions, $C_i(x_1, x_2)$, lie at the boundary of the set.

Assume that we have another set $\{D_j\}$, where the elements are also basic two-electron densities. We furthermore assume that every element in $\{D_i\}$, can be written as a sum of elements in $\{C_i\}$, with positive coefficients, i. e.:

$$\mathbf{D} = \mathbf{A}\,\mathbf{C} \tag{15}$$

where the elements of A are all positive or zero. We then have:

$$\mathbf{C} = \mathbf{A}^{-1}\,\mathbf{D} \tag{16}$$

since the determinant of A is not zero. If it were, the elements of D would be linearly dependent. It is obvious, however, that some of the elements of $A^{-1}$ must be negative unless A is a unit matrix (for some ordering of the elements of C). Thus there are elements in $\{C_i\}$, that cannot be expressed in $\{D_i\}$, using positive coefficients, contradicting the assumption that $\{D\}$ consists of basic two-electron densities. We conclude that the set $\{C_i\}$, is unique up to ordering.

We conjecture that the set of basic two-electron densities is unique also when the cardinality of $\phi$ is infinite. In this case, $\phi$ will define a convex set in an infinite-dimensional Hilbert space instead of a convex set in a finite vector space.

Concerning the two-electron density matrix $\Gamma(x_1', x_2' | x_1, x_2)$ we see from Eqn. (5) that it is "in some sense" only the diagonal which is relevant. With this we mean that if we know $\Gamma(x_1', x_2' | x_1, x_2)$ when $|x_1' - x_1| < \varepsilon$ and $|x_2' - x_2| < \varepsilon$ for arbitrary small positive $\varepsilon$ then we can evaluate Eqn. (5). It is however not possible to replace the primed variables with the unprimed ones, since they are independent.

A valid two-electron density matrix is given by:

$$\Gamma_i(x_1',x_2'|x_1,x_2) = \binom{N}{2}\int \overline{\phi_i(x_1',x_2',...,x_N)}\phi_i(x_1,x_2,...,x_N)dx_3...dx_N \tag{17}$$

By the same reasoning as for the two-electron densities, we see that we can define a set of basic two-electron density matrices. If we have a normalized sum of two proper basis functions: $\varphi = \alpha_i \phi_i + \alpha_k \phi_k$ and form the density matrix we find:

$$\Gamma_\varphi(x_1',x_2'|x_1,x_2) = \binom{N}{2} \int \overline{(\alpha_i\phi_i + \alpha_k\phi_k)}(\alpha_i\phi_i + \alpha_k\phi_k)dx_3...dx_N =$$

$$= |\alpha_i|^2 \Gamma_i(x_1',x_2'|x_1,x_2) + |\alpha_k|^2 \Gamma_k(x_1',x_2'|x_1,x_2) + \tag{18}$$

$$\binom{N}{2} \int \overline{\alpha_i\phi_i(x_1',x_2',x_3,...)}\alpha_k\phi_k(x_1,x_2,x_3,...) + \overline{\alpha_k\phi_k(x_1',x_2',x_3,...)}\alpha_i\phi_i(x_1,x_2,x_3,...)dx_3...dx_N$$

which is similar to Eqn. (11). The last term is zero on the diagonal, but not necessarily for the off-diagonal elements. It may be expected, that by choosing a sufficiently small stripe on the diagonal, i. e., a small $\varepsilon$, that we can ignore the off-diagonal elements. With sufficiently small $\varepsilon$, we mean that $\varepsilon < a/2$ where a is the lattice constant of (or distance between) the basis functions $\chi_i$. This means, in other words, that the last term in Eqn. (18) is zero. We will however show that it is necessary to take into account the non-diagonal terms in Eqn. (18).

We have an approximation problem. The problem is that the basis functions we use are very badly suited for computing differentials, having zero derivatives almost everywhere. If the number of proper basis functions is finite (as is always true in practice) it will be necessary to use differences instead of differentials, when minimizing Eqn. (5). This means that we cannot use $|x_1'-x_1| < \varepsilon$ and $|x_2'-x_2| < \varepsilon$, with a very small $\varepsilon$, in the two-particle density matrix as the differences will then always be zero, which is nonsense. We must include the off-diagonal (interference) terms in Eqn. (18) and the set of two-particle density matrices will not form a convex set with the basic density matrices at the boundary, in contrast to the two-electron density.

A different choice of $\chi_0$ can be made such that Eqn. (7) is satisfied and such that the derivatives of $\chi_0$ are non-zero. For example $\chi_0$ could be a Gaussian within the hypercube, and such a set of functions can be made dense. In such a case there will be a simple and fixed relationship between $\chi_0$ and $\nabla\chi_0$ such that $\nabla\Gamma(x_1', x_2'|x_1, x_2)$ will not be independent of $\Gamma(x_1', x_2'|x_1, x_2)$ for $|x_1'-x_1|<\varepsilon<a/2$ and $|x_2'-x_2|<\varepsilon<a/2$ (with $a$ the distance between the functions $\chi_i$) which we require.

If the density matrices did form a convex set the minimum of Eqn. (5) would occur at the boundary of this set, which in a sense seems a "too easy" solution for the few-body problem.

From these results we can suggest a new variational method for finding the ground state energy (and two-electron density) of the few-body problem. We first construct a set of two-electron densities $C_i(x_1, x_2)$ using any set of suitable wave-functions, not necessarily proper. The set of wave-functions chosen is determined by the user and their number must not necessarily be very high. From these two-electron densities we can obtain new valid two-electron densities using the fact that all N-representable two-electron densities lie in a convex set. We can for such two-electron densities evaluate the exchange and correlation energies exactly. The energy from an external potential can also easily be found. We then have to evaluate the kinetic part of the energy and we here have to resort to some approximate functional of the two-electron density. This could be a local one-electron density approximation, using data from the homogeneous electron gas, which can then be extended by gradient expansions, or more general types of functionals[12]. It is important to note that we only have to find approximations to the kinetic part, which hopefully can be done more accurately than to find approximations for the total energy, as is done in density functional theory. The scaling properties of the kinetic-energy functional are for instance much simpler than for the total energy. To find accurate functionals for the kinetic energy is not easy but work is being done in the formulation of functionals for the kinetic energy based on the one-electron density[13] which avoid the Kohn-Sham equations. Given the approximate kinetic energy and the exact correlation and external energies, we then minimize the total energy over the N-represesentable two-electron densities to find the ground state energy. By minimising over N-represesentable two-electron densities we avoid issues of v-representability.

In summary, we have found that the two-electron densities lie in a convex set and we have determined the boundary of this set. Using this information we suggest a new variational method to solve the few-particle problem, where the exchange-correlation part of the energy is treated exactly.

This work was performed within the nanometer consortium in Lund and supported in part by VR, SSF and in part by the European Community's Human Potential Program under contract HPRN-CT-2002-00298. We acknowledge clarifying discussions with Drs. C.-O. Almbladh, U. von Barth, P. Kurasov, and S. Silvestrov.